# Convolutional neural network-based single-shot speckle tracking for x-ray phase-contrast imaging

Serena Qinyun Z. Shi, Nadav Shapira, Peter B. Noël, *Member, IEEE*, and Sebastian Meyer

*Abstract*—**X-ray phase-contrast imaging offers enhanced sensitivity for weakly-attenuating materials, such as breast and brain tissue, but has yet to be widely implemented clinically due to high coherence requirements and expensive x-ray optics. Speckle-based phase contrast imaging has been proposed as an affordable and simple alternative; however, obtaining high-quality phase-contrast images requires accurate tracking of sample-induced speckle pattern modulations. This study introduced a convolutional neural network to accurately retrieve sub-pixel displacement fields from pairs of reference (i.e., without sample) and sample images for speckle tracking. Speckle patterns were generated utilizing an in-house wave-optical simulation tool. These images were then randomly deformed and attenuated to generate training and testing datasets. The performance of the model was evaluated and compared against conventional speckle tracking algorithms: zero-normalized cross-correlation and unified modulated pattern analysis. We demonstrate improved accuracy (1.7 times better than conventional speckle tracking), bias (2.6 times), and spatial resolution (2.3 times), as well as noise robustness, window size independence, and computational efficiency. In addition, the model was validated with a simulated geometric phantom. Thus, in this study, we propose a novel convolutional-neural-network-based speckle-tracking method with enhanced performance and robustness that offers improved alternative tracking while further expanding the potential applications of speckle-based phase contrast imaging.**

## I. INTRODUCTION

X-ray phase-contrast imaging (PCI) has proven to be a powerful technique for non-destructive material testing and biomedical imaging [1], [2]. While conventional x-ray imaging relies on absorption in high-density materials for signal generation, PCI measures changes in the wavefront (phase shift) when x-rays pass through an object. For typical x-ray energies and materials of low atomic numbers - such as human tissue - the generation of phase shift is several orders of magnitude larger than absorption. Therefore, for weakly-attenuating materials, PCI provides enhanced sensitivity (i.e., visualization of soft-tissue contrast) that is inaccessible in conventional x-ray imaging. The clinical potential of PCI has been demonstrated for a wide range of pathologies and anatomical sites, such as the musculoskeletal system [3], central nervous system [4], breast [5], and vasculature [6].

Although various solutions for sensing x-ray phase information have been developed in the last decades [7], their widespread clinical application is still limited to prototypes [8]–[10]. Typical PCI systems utilize 1D or 2D gratings (grating interferometry [11], [12]) and grids to produce a periodic reference interference pattern in the detector plane. However, the translation of these PCI systems from research laboratories to clinical centers faces major obstacles because of high coherence requirements, complex optical systems for translation of phase shifts into measurable intensity variations, and phase-wrapping effects from periodic reference patterns [13].

Speckle-based x-ray phase contrast imaging (XPCI) [13]–[16] is a recently proposed method for phase-contrast and dark-field imaging that utilizes x-ray near-field speckles generated from a random diffuser. The principle of XPCI is schematically shown in Fig. 1. Coherent x-rays impinge on the diffuser, randomly scatter, and mutually interfere with the incident beam to create a random intensity pattern, named the reference image. Sample-induced phase shifts cause refraction that translates into a transverse displacement of the original speckle pattern to generate the sample image. The sample image can then be compared to the reference to calculate the corresponding phase contrast signal of the object [13]. XPCI overcomes several of the limitations of typical PCI systems as it offers a simple setup with excellent dose efficiency, only has moderate coherence requirements, does not require precise system alignment, and negates the propagation distance restrictions imposed by fractional Talbot distances [15], [17]. This is crucial for preclinical PCI systems, potentially used for small animal imaging, due to less

This paper was first submitted on April 30, 2023.

S. Q. Z. Shi is with the Department of Radiology, Perelman School of Medicine and Department of Bioengineering, University of Pennsylvania, Philadelphia, PA 19103, USA (e-mail: sqzshi@seas.upenn.edu).

N. Shapira is with the Department of Radiology, Perelman School of Medicine, University of Pennsylvania, Philadelphia, PA 19103, USA (e-mail: nadavshap@gmail.com).

P. B. Noël is with the Department of Radiology, Perelman School of Medicine, University of Pennsylvania, Philadelphia, PA 19103, USA (e-mail: peter.noel@pennmedicine.upenn.edu).

S. Meyer was with the Department of Radiology, Perelman School of Medicine University of Pennsylvania, Philadelphia, PA 19103 USA. He is now with Memorial Sloan Kettering Cancer Center, New York, NY, 10065 USA (e-mail: meyers4@mskcc.org).

Research reported in this publication was supported by the National Institutes of Health (NIH) (R01HL166236, 5T32EB009384-12) and the National Science Foundation (NSF) (DGE-1845298). The content is solely the responsibility of the authors and does not necessarily represent the official views of the NIH, ITMAT, or NSF. We thank Leening P. Liu (University of Pennsylvania) for comments on the manuscript.

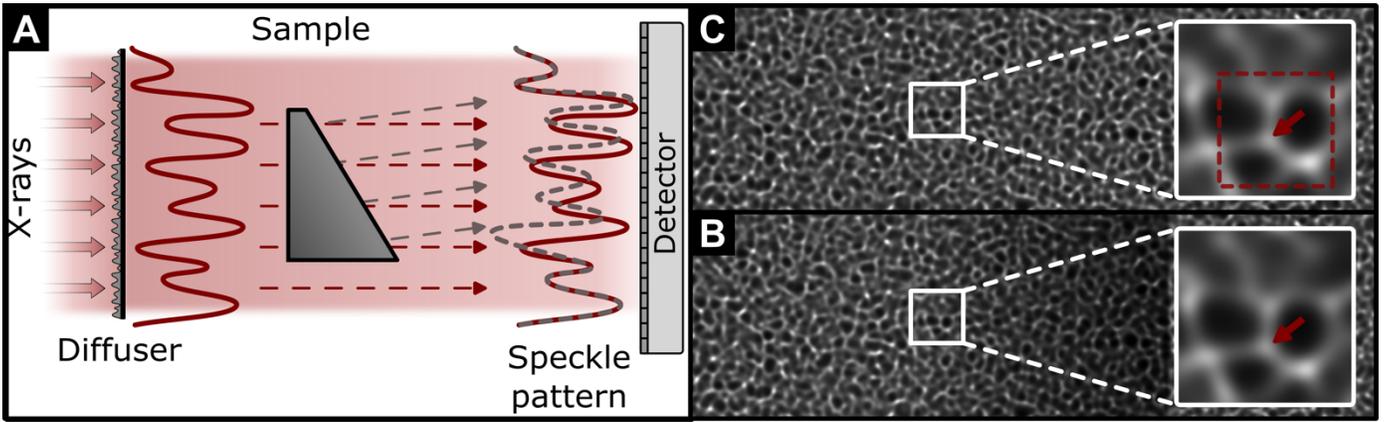

Figure 1. Principle of speckle-based phase-contrast imaging. **(A)** A random intensity pattern (red solid line) is shown in comparison to the displaced intensity pattern in the sample image (gray dashed line). **(B)** and **(C)** show a reference and sample image, respectively, and demonstrate a sub-pixel displacement of the speckle marked by the red arrow.

stringent requirements for small detector pixels and long propagation distances compared to laboratory setups. The key to obtaining high-quality phase contrast images from XPCI systems is accurate tracking of the sample-induced speckle pattern modulations. Out of all speckle tracking modes [13], single-shot speckle tracking (XST), i.e., using only one reference and sample image pair, is desirable for a preclinical translation since it allows a fast and dose efficient acquisition with a stationary diffuser. Several algorithms have been successfully developed for XST. Zero-normalized cross-correlation (ZNCC) [14] and unified modulated pattern analysis (UMPA) [16], [18]–[20] are direct tracking algorithms based on windowed image correlation. Although both algorithms produced impressive results, the trade-off between spatial resolution and angular sensitivity requires careful selection of the window size [19]. One possible solution to overcome this limitation is the optical flow method (OF) [21]. By implicitly tracking speckles, i.e., without the use of a correlation window, measuring displacement fields can be considered an optical flow problem through geometrical-flow conservation. A study by Rougé-Labriet *et al.* [22] established that of the three speckle tracking techniques, the OF method provided the best qualitative image quality and the lowest naturalness image quality evaluator score with a reduced number of sample exposures for low dose PCI with both theoretical and biomedical sample models. However, this method depends on the assumption that the sample is transparent to x-rays and utilizes a high-pass filter which can result in image artifacts and affect the quantitative accuracy.

Convolutional neural network (CNN) have been successfully implemented for various problems in computer vision [23], focusing on classification [24], segmentation [25], and registration [26]. More recently, CNNs have been extended to the general optical flow problem, defined as the pattern of apparent motion of objects between two frames, using deep learning architectures [27], [28]. FlowNet [29] and its variants [27] use the multiscale loss function for optimization and are U-shaped with contracting and expanding paths. FlowNet2 [28], a fusion network generated by stacking different FlowNet variants, achieved superior performance compared to traditional optical flow algorithms. This architecture has been successfully utilized for displacement estimation in ultrasound elastography [27] and various applications in civil engineering [30]. However, compared to these applications, the sample-induced displacement for x-ray speckle tracking is typically much smaller at subpixel levels. The StrainNet architecture, designed by Boukhtache *et al.* [23], can retrieve dense displacement and strain fields from optical images of an object exposed to mechanical compression. StrainNet has successfully demonstrated comparable results in retrieval performance and computing time compared to traditional algorithms. With its ability to perform subpixel displacement retrievals, StrainNet could be a promising solution for XST.

In this paper, we present the CNN-based Analysis for Displacement Estimation (CADE), an extension of the StrainNet CNN algorithm to track x-ray speckles in XPCI. Intrinsic performance characteristics for CADE were quantitatively investigated using standard criteria for digital image correlation and compared to established x-ray speckle tracking algorithms. In addition, the performance of CADE for speckle-based PCI was evaluated using numerical wave-optics simulations.

## II. METHODS

### A. Wave-optics simulation

Numerical wave-optics simulations were performed using a previously-developed in-house Python simulation framework [31]. The simulation process relied on an iterative use of the angular spectrum method to propagate the wave-field from the source through the speckle-based imaging setup. The disturbance of the wave-field due to the presence of an object (i.e., attenuation and phase-shift) was then calculated in projection approximation. All simulations were conducted with the following configuration. A monochromatic 30 keV point source with a 10 μm focal spot was simulated. The diffuser was located 1 m away from the source and was modeled as 10 layers of sandpaper sheets, each consisting of a rough aluminum oxide ($Al_2O_3$) surface with a 200 μm backing of diethyl pyrocarbonate ($C_6H_{10}O_5$)

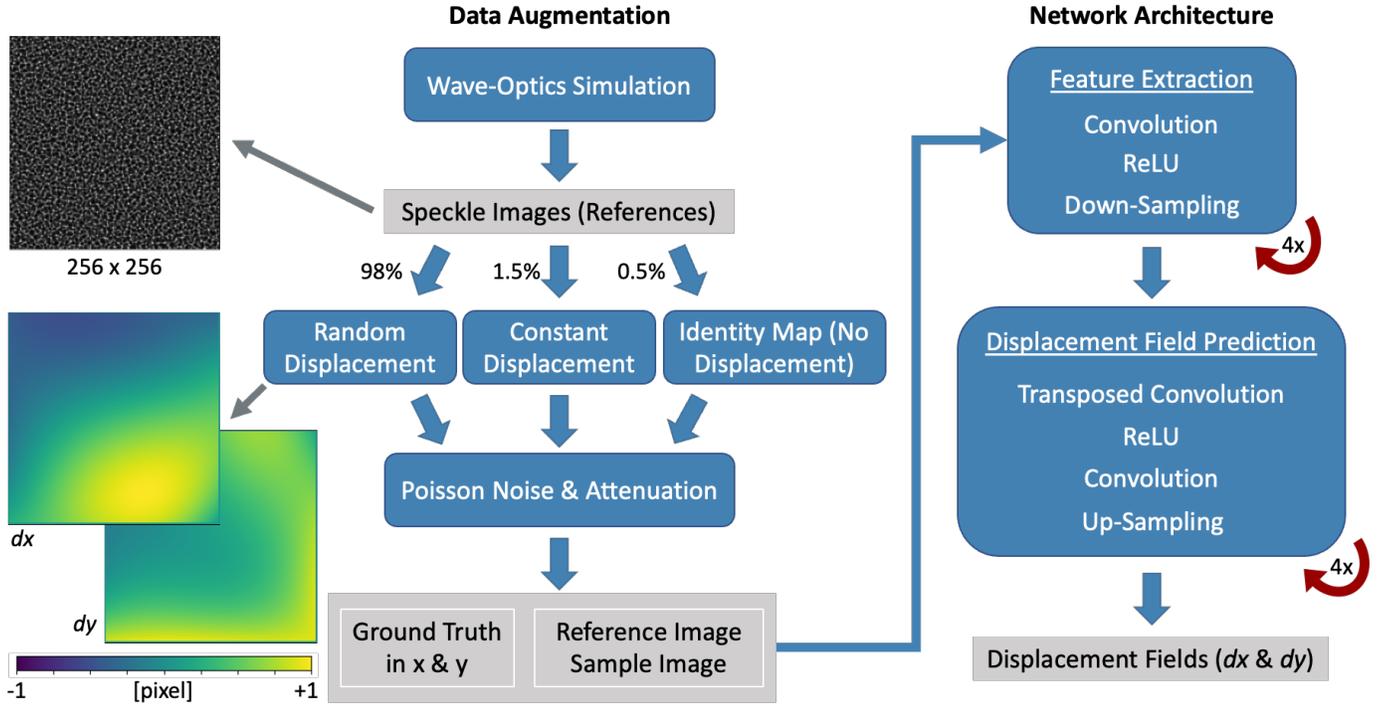

Figure 2. Schematic view of data augmentation and network architecture. Percentages show the proportion of data that underwent the respective processing. Examples of reference and displacement images are shown on the left (gray arrows). The feature extraction level and displacement field prediction level were both performed four times. ReLU stands for rectified linear unit. Down-sampling and up-samplings were performed at a stride of 2.

[31]. The detector was located 3 m away from the source and had an effective pixel size of 12 μm, utilizing a point spread function of 1/2.355 pixels. As the diffuser is simulated with different surface structures, the resulting speckle sizes represented by the full width half maximum (FWHM) of the speckle pattern autocorrelation function ranged from 22 μm to 110 μm, or approximately 2 to 10 pixels at the detector level. This range offered a minimum speckle visibility of 20% and a good representation of expected speckle sizes.

*B. Data augmentation*

Data augmentation for supervised training and network architecture details are shown in Fig. 2. As described in Section II-A, wave-optics simulations were utilized to obtain 364 independent 256 x 256 pixel reference speckle images. Random piece-wise smooth deformations were applied to each reference image using one of six deformation patch sizes (4, 8, 16, 32, 64, or 128 pixels) with displacements ranging from −1 to +1 pixel. The deformation patch size indicates the distance of linear interpolation of displacements, i.e., a patch size of 8 x 8 pixels corresponded to independent 8 x 8 patches of smooth displacements in only one direction (either all positive or all negative). The random deformations were applied to each reference image to generate the corresponding sample image. Sixty and 10 independent deformations were used for each reference image to generate the network training and testing datasets, respectively. The generated displacement values follow a normal distribution centered around 0 and ranging from -1 to +1 pixels.

We randomly selected 0.5% of all image pairs to utilize each of the following deformation maps: (1) identity maps (i.e., all displacements equal zero); (2) constant displacement (i.e., the same displacement value across the entire map) in x direction and an identity map in y direction; (3) vice versa of (2); and, finally, (4) constant displacements in both x and y directions. This resulted in 98% of the data having random deformations, and the remaining 2% underwent identity or constant displacement maps. The constant displacements within ±0.15 pixels were included to improve CADE's performance at extremely small displacements.

Additional processing of image pairs included noise and attenuation. First, individual Poisson noise maps were generated and applied for reference and sample images. All sample images were then randomly attenuated to mimic 50 – 100% transmission in the same manner as the deformation patches. This resulted in a training and testing dataset of 21841 and 3640 image sets, respectively. Each data set comprises a reference image, sample image, and ground truth displacement field.

*C. CNN architecture and training*

*1) Network architecture*: The StrainNet-f architecture [23] adapted for XPCI (Fig. 2) is an end-to-end full-resolution network consisting of two main components. The first component extracted feature maps with successive convolutional layers. The 10 convolutional layers include 7 x 7 filters for the first, 5 x 5 filters for the second and third, and 3 x 3 filters for the remaining seven. The latter portion predicts displacement fields via five convolutional layers with 3 x 3 filters and eight transposed convolutional layers. The architecture simplified FlowNetS with four down-samplings and four up-sampling. The same loss function and levels in FlowNetS were used.

Table 1. Network hyperparameters used for training. The highlighted parameters deviated from the initial configuration. Beta corresponds to the beta parameter for the Adam solver algorithm. Div Flow represents the value by which the flow will be divided every 40 epochs to decrease the runtime.

| Training Hyperparameters | | | | | |
|---|---|---|---|---|---|
| Bias Decay | 0 | Epoch Size | 0 | Weight Decay | 0.0004 |
| Solver Algorithm | Adam | Batch Size | 16 | Algorithm | StrainNet_f |
| Div Flow | 2 | Learning Rate | 0.001 | Multiscale Weights | [0.005, 0.01, 0.02, 0.08, 0.32] |
| Epochs | 350 | Momentum | 0.9 | Data Loading Workers | 8 |
| Starting Epoch | 0 | Beta | 0.999 | Milestones | [40, 80, 120, 160, 200, 240] |

*2) CNN training:* The hyperparameters of the network were initially set to the original StrainNet-f configuration [23] and further fine-tuned via grid search to maintain equal or improved model convergence and faster training times. The final values of each hyperparameter are reported in Table 1.

Model convergence was evaluated with training and testing endpoint error (EPE), which is the Euclidean distance between the predicted and ground truth displacement vectors normalized over all pixels:

$$L_{epe} = \frac{1}{N} \sum \sqrt{(\mathbf{u}_{GT}(x,y) - \mathbf{u}(x,y))^2}, (1)$$

where $N$ denotes the total number of pixels in the image, $\mathbf{u}_{GT}(x,y)$ is the ground truth displacement of each pixel, and $\mathbf{u}(x,y)$ is the estimated displacement of pixel $x, y$ [23]. The training was performed with four cores of an NVIDIA Tesla T4 16GB GPU at a runtime of 58 hours. An EPE of 0.050 for training and 0.113 for testing was achieved after 350 epochs.

### D. State-of-the-art speckle tracking algorithms

To reconstruct the (differential) phase-contrast image, the displacement vector field ($\mathbf{u}(x,y)$) between the sample ($\mathbf{I_s}$) and reference ($\mathbf{I_r}$) image must be determined by locally tracking the speckle pattern modulation. Two conventional speckle tracking algorithms were examined: zero-normalized cross-correlation (ZNCC) [32] and unified modulated pattern analysis (UMPA) [18], [19], [33].

*1) ZNCC:* Small patches of size $(2M + 1) \times (2M + 1)$ from $\mathbf{I_r}$ were compared against a template region in $\mathbf{I_s}$. The relative transverse displacement for the central pixel $(x_0, y_0)$ of the template was determined as the shift with the highest correlation coefficient:

$$\mathbf{u}(x_0, y_0) = argmax_{u_x, u_y} \left\{ \frac{\sum_{i,j}[\mathbf{I'_s}(x_i, y_j)\mathbf{I'_r}(x_i + u_x, y_j + u_y)]}{\sqrt{\sum_{i,j} \mathbf{I'_s}(x_i, y_j)^2 \sum_{i,j} \mathbf{I'_r}(x_i + u_x, y_j + u_y)^2}} \right\}, (2)$$

where $\mathbf{I_r'}$ and $\mathbf{I_s'}$ are the normalized reference and sample images obtained by subtracting the mean value of the patch. The summation was performed over all pixels in the corresponding patch. Sub-pixel precision was obtained by Gaussian fitting to the peak in the correlation map.

*2) UMPA:* A physical model is used to describe the influence of the sample on the speckle pattern in terms of transmission T and transverse speckle displacements $(u_x, u_y)$. All signals are extracted with a windowed least-square minimization between the model and the measured sample image $I_s$:

$$\mathbf{u}(x_0, y_0) = argmin_{u_x, u_y} \sum_{i,j} w(x_i, y_j) \times \{I_s(x_i, y_j) - T(x_i, y_j)I_r(x_i, y_j)I_r(x_i + u_x, y_j + u_y)\}^2, (3)$$

where $w$ is a windowing function of $(2M + 1) \times (2M + 1)$ pixels centered on $(x_0, y_0)$. Sub-pixel precision was achieved through a paraboloid fit of the neighborhood of the minimum.

### E. Performance validation and comparison

Ten independent reference images were generated with speckle parameters described in Section II-B and used for validation and comparison evaluations. Different types of displacement maps were applied for each evaluation. Poisson noise and transmission of 90% were applied to all image pairs unless otherwise stated. All algorithms were run on a MacBook Pro with an Apple M1 chip.

*1) Speckle tracking accuracy:* Before adding noise and attenuation, each reference image was transformed with a constant displacement map ranging from 0 to 1 pixel at steps of 0.1 pixels. Bias and root mean squared difference (RMSE) of the retrieved displacement fields were calculated as [34], [35],

$$Bias(x,y) = \left(\mathbf{u}_{GT}(x,y) - \frac{\sum \mathbf{u}(x,y)}{N}\right) \div \mathbf{u}_{GT}(x,y), (4),$$

$$RMSE(x,y) = \sqrt{\frac{\sum_{x,y=1}^{N}(\mathbf{u}(x,y) - \mathbf{u}_{GT}(x,y))^2}{N}}, (5).$$

*2) Spatial resolution:* The spatial resolution of each algorithm was evaluated using a star displacement map, which consisted of a unidirectional sinusoidal displacement with linearly increasing frequency toward the left. A key characteristic of the pattern is the constant amplitude of 0.5 pixel across the horizontal symmetry axis in the center of the image. Limiting spatial resolution for each algorithm was defined by the frequency at a bias of 10% [23].

*3) Noise, window size dependency, and computational time:* Reference images were deformed by a gradient map ranging from 0 to +1 pixel displacement. The effect of noise was then evaluated by applying seven different noise levels ranging from a loss of 0-6% signal-to-noise ratio (SNR) in the same manner as in Section II-B. RMSE and spatial resolution were calculated for UMPA and ZNCC with window sizes between 10 and 50 pixels and compared to CADE to examine window size dependency. Finally, computational times were evaluated using 10 image pairs of 256 x 256, 512 x 512, and 768 x 768 pixels.

*4) Method validation:* The wave-optics simulation was used to validate our speckle tracking method for imaging data obtained from an XPCI acquisition of with the setup described in Section II-A. The simulated polymethyl methacrylate object consisted of a 1500 μm wide rectangular base with a thickness profile modulated in the x direction by a sine wave. Hence, the thickness $t(x, y)$ of the sample is represented by

$$t(x,y) = b + A \cdot \sin(wx), (6)$$

where $b = 1000$ μm is the base thickness of the sample and $A = 800$ μm and $w = 1.33 \times 10^{-3}$ μm$^{-1}$ are the amplitude and frequency of sinusoidal thickness modulation, respectively.

The gradient of the phase shift Φ introduced by the sample is proportional to the refraction angle $\alpha = (\alpha_x, \alpha_y)$, where $x$ and $y$ are the transverse coordinates orthogonal to the optical axis. This can in turn be geometrically related to the speckle displacement vector $\mathbf{u}(x,y) = (u_x(x,y), u_y(x,y))$ (see Fig. 1) in small-angle approximation:

$$\left(\frac{\partial \Phi}{\partial x}, \frac{\partial \Phi}{\partial y}\right) = \frac{2\pi}{\lambda}(\alpha_x, \alpha_y) = \frac{2\pi}{\lambda}(u_x, u_y)\frac{p}{d} \quad (7)$$

where $\lambda$ is the x-ray wavelength, $p$ is the detector pixel pitch, and $d$ is the sample-detector distance [36]. The phase shift can also be calculated from the sample properties as

$$\Phi(x,y) = -\frac{2\pi\delta}{\lambda}t(x,y) \ (8)$$

using the refractive index decrement $\delta$ of the sample's material. Hence, the ground truth refraction angle $\alpha_x(x,y)$ for this sample is given by

$$\alpha_x(x,y) = \frac{\lambda}{2\pi}\frac{\partial \Phi(x,y)}{\partial x} = -\delta\frac{\partial t(x,y)}{\partial x} = -\delta Aw \cdot \cos(wx). (9)$$

## III. RESULTS

CADE improved speckle tracking accuracy in terms of RMSE and bias for almost all displacement values (see Figure 3). The mean RMSE was $14 \times 10^{-3}$, $24 \times 10^{-3}$, and $22 \times 10^{-3}$ pixels, while mean bias values were $7 \times 10^{-3}$, $16 \times 10^{-3}$, and $18 \times 10^{-3}$ pixels for CADE, UMPA, and ZNCC, respectively. Spatial resolution was visibly improved (see Fig. 4) with CADE, particularly in the high frequency regions. Quantitatively, CADE provided the best spatial resolution of 56 pixel in comparison to 68 pixel and 126 pixel for the ZNCC and UMPA algorithms, respectively.

With varying amounts of noise, CADE performed better than ZNCC. For our configuration, UMPA performance rapidly degraded for increasing noise levels, resulting in substantially higher RMSE values. Noting that CADE does not require a window, Fig. 5 shows that CADE obtained superior accuracy and spatial resolution compared to ZNCC and UMPA, except for very small window sizes below 15 pixel, which are associated with low spatial resolution. All three algorithms showed a linear relationship between computational time and image size. CADE resulted in shorter computational times, and its advantage increased substantially with larger image sizes. At an image size of around $6 \times 10^5$ pixels, CADE had an average runtime of 18 s, whereas the runtimes for ZNCC and UMPA were three (53 s) and ten times (181 s) longer, respectively.

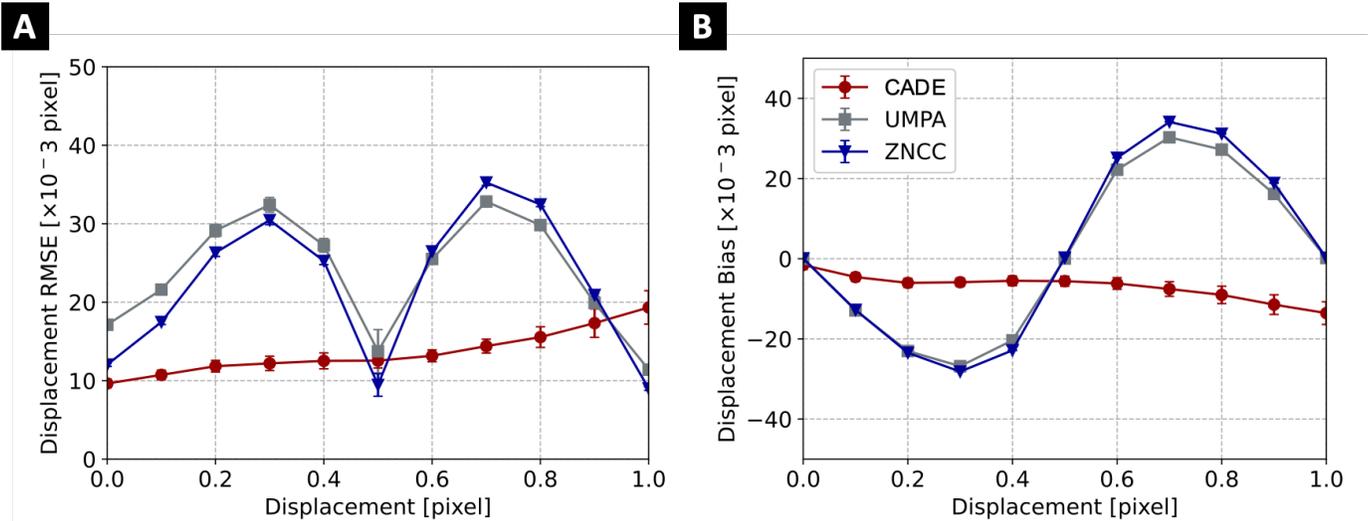

Figure 3. Displacement RMSE (**A**) and bias (**B**) for constant displacement maps ranging from 0 to 1 pixel for CADE, UMPA, and ZNCC speckle tracking algorithms. Mean (central marker) and standard deviation (error bar) were obtained from 10 validation image pairs for each displacement value.

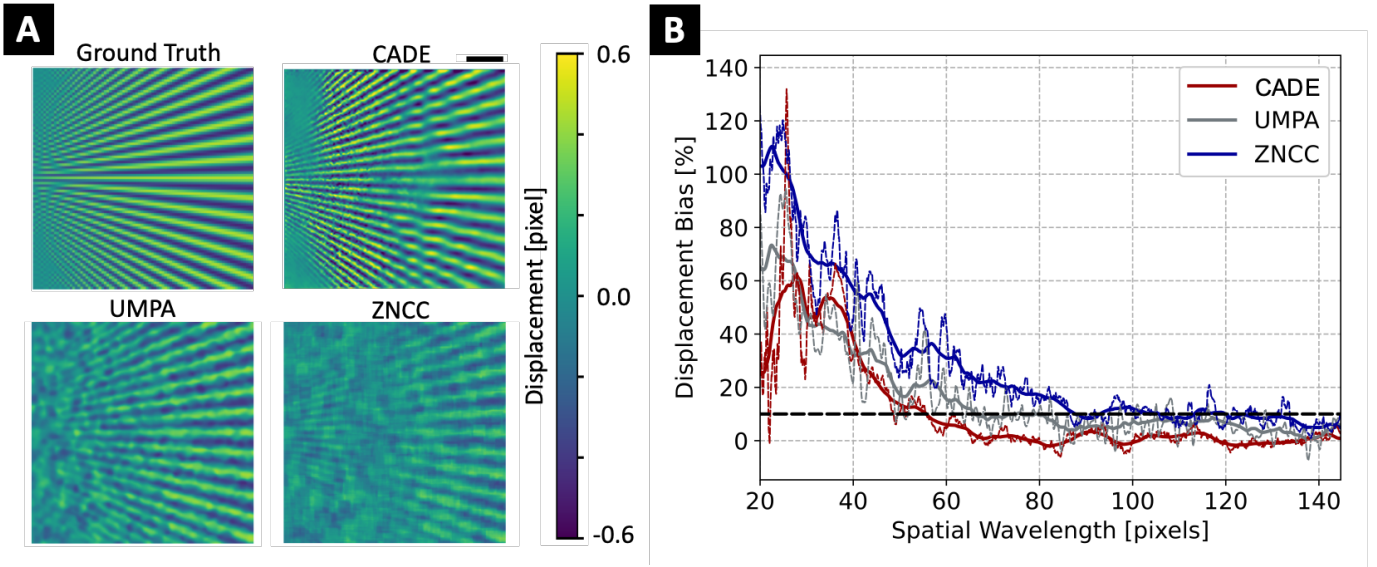

Fig. 4. Evaluation of the speckle tracking spatial resolution with star pattern displacement maps. (**A**) High frequency area of the star pattern and corresponding displacement maps obtained from the speckle tracking algorithms. Scalebar represents 0.1 mm. (**B**) Spatial resolution of CADE and conventional algorithms. True (thin dotted line) and moving averaged (thick solid line) displacement bias of each method is shown versus the spatial wavelength of the star pattern. Spatial resolution is defined by 10% bias (black dotted line).

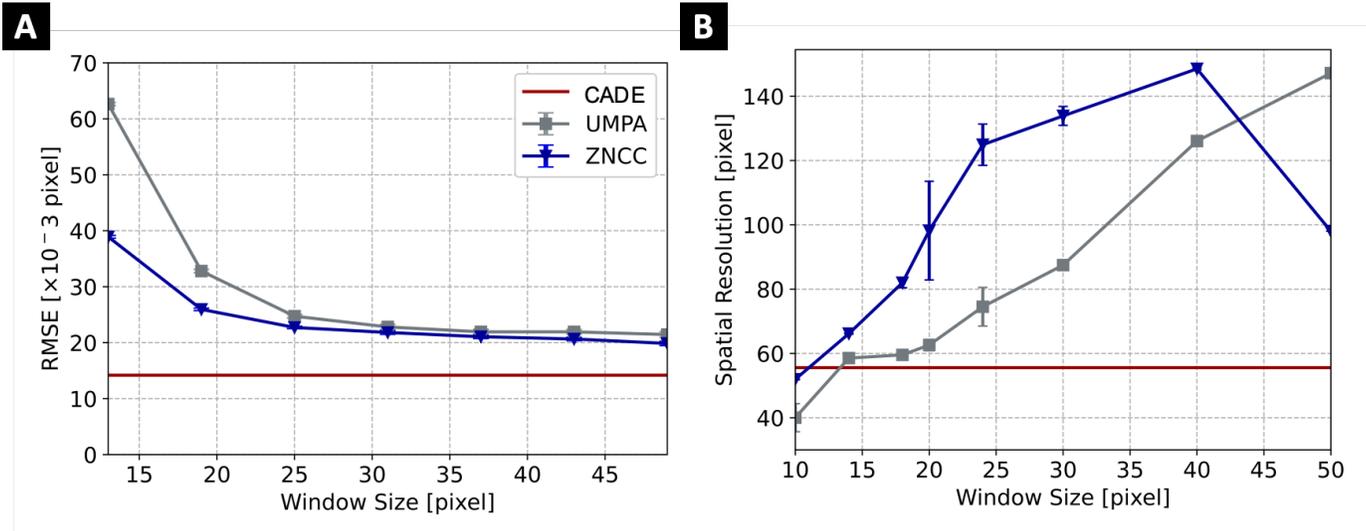

Figure 5. Comparison of displacement RMSE (**A**) and spatial resolution (**B**) of conventional algorithms for increasing correlation window size. The CADE results do not dependent on window size. Mean (marker) and standard deviation (error bar) are obtained from evaluating ten image pairs for each window size. CADE achieved improved RMSE for all window sizes and improved spatial resolution for window sizes greater than 15 pixels.

Displacement maps and reconstructed refraction angle for the sine wave sample are shown in Fig. 6. Qualitatively, CADE-generated displacement maps appeared less noisy and in better agreement with the ground truth than for conventional algorithms. In Fig. 6B, CADE achieved improved accuracy compared to ZNCC and UMPA, particularly at the peaks located at ±2 μrad. The displacement maps RMSE was 4.10, 4.37, and $4.30 \times 10^{-2}$ pixels for CADE, UMPA, and ZNCC, respectively.

## IV. DISCUSSION

XPCI presents a cost-effective method with moderate coherence requirements for enhanced sensitivity compared to conventional x-ray systems. However, accurately tracking sample-induced speckle pattern modulations is crucial for obtaining high-quality phase contrast images. Although several methods have been proposed for this task, they often necessitate a trade-off between tracking accuracy and spatial resolution or rely on several assumptions. To overcome these limitations, we present CADE, a novel windowless CNN-based speckle tracking algorithm, and compare and validate its performance against conventional algorithms. The key findings from this study are: (1) successful application of CADE for speckle tracking, (2) improved tracking performance compared to conventional algorithms, and (3) greatly reduced computational time. Most importantly, CADE achieved superior performance particularly at high refraction angles when validated on a simulated object.

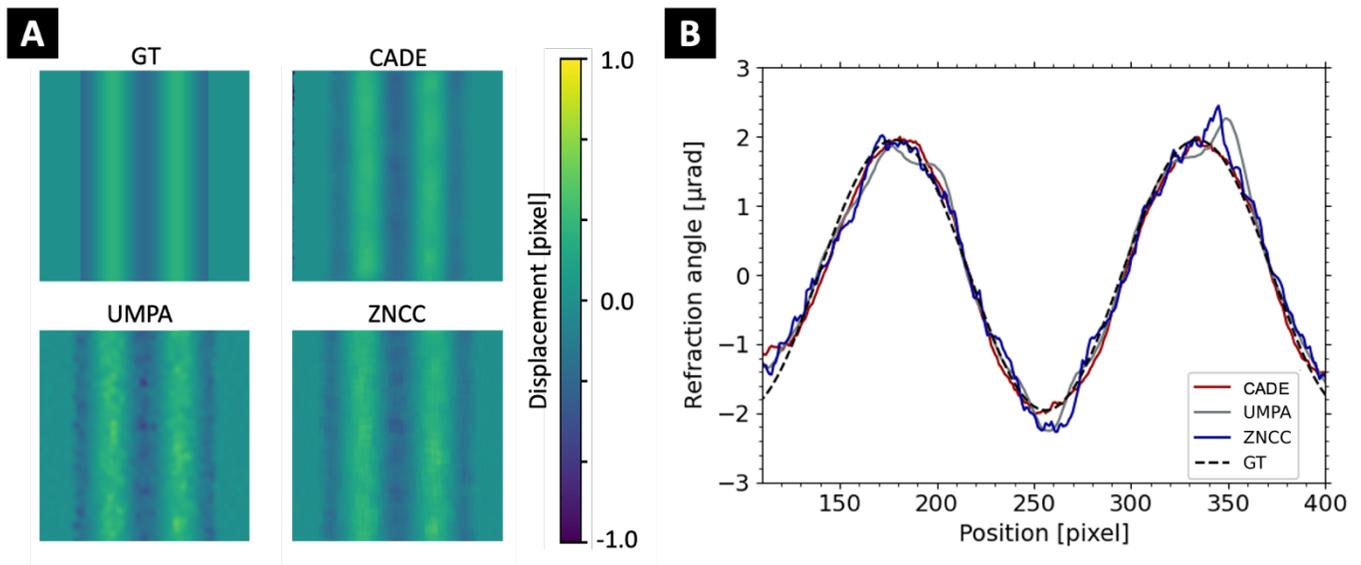

**Figure 6.** (**A**) Displacement maps of ground truth (GT), CADE model, ZNCC, and UMPA algorithms for the XPCI wave-optics simulation of the geometric phantom . (**B**) Corresponding refraction angle profiles obtained.

Compared to CADE, current state-of-the-art algorithms like UMPA and ZNCC are window-based algorithms. Unlike these extrinsic approaches, or iterative pixel-wise algorithms, intrinsic speckle tracking algorithms rely on solving a partial differential equation formulated at the whole-image level rather than explicitly tracking individual speckles. Recent examples include the geometric-flow approach [21] and multimodal intrinsic speckle-tracking [37], which combines the geometric-flow formalism with a Fokker-Planck-type generalization. A recent publication by De Marco's [33] presents an enhanced implementation of UMPA, characterized by greatly improved computation efficiency, the capability of multithreading, and the reduction of estimation bias. As we implemented an older version of UMPA, the performance of this updated UMPA algorithm is unknown, but we believe that the general trend is similar to what we have presented in this paper. Finally, a machine learning method for speckle tracking with model validation on experimental data has been proposed lately [38]. While both algorithms utilized simulated random displacement maps for training and offered improved runtime and image quality, there were several distinct differences: (a) speckle patterns were generated with a coded binary mask versus our random sandpaper model; (b) they used a basic plane-wave model while we utilized a divergent-beam geometry for image propagation; (c) noise was calculated with either a random binary noise image or as Gaussian noise whereas we used Poisson noise; and, (d) their model was based on the SPINNet architecture while ours is adapted from the StrainNet-f architecture. We believe that our approach is a more realistic solution for speckle tracking as it utilizes more representative training and testing data from the sandpaper model and wave-optics simulation. On the other hand, further evaluations are needed to assess and compare the performances.

Although CADE demonstrates improved speckle tracking performance and overcomes several of the issues of conventional algorithms, it still suffers from limitations. For example, CADE exhibits a slight decrease in accuracy for increasing displacements, which may be explained by the training data distribution (centered around zero and ranging from -1 to +1 pixel) due to the patch-based deformation method. However, as the intended application of this study focuses on the retrieval of subpixel displacements, CADE provides superior accuracy and spatial resolution compared to cross-correlation methods. The main limitation of our study is the lack of experimental image data for both training and validation purposes. Speckle patterns can be very well characterized by their statistical properties and thus simulation and numerical data can be reliably used to model a realistic diffuser setup. Although we were able to successfully validate CADE on a simulated geometric sample, training and testing with real image data with more complex samples would be ideal as this would produce the most realistic model but is unrealistic due to the large amount of data required by model training (i.e., currently requiring datasets of 25481 sets of images).

In conclusion, this study successfully implemented and validated CADE, a windowless CNN-based speckle tracking method, which demonstrated superior performance, greatly decreased processing times, and robustness to noise. Furthermore, due to its ability to process much higher volumes of data with equal or better tracking accuracy, CADE brings the development and application of single-shot, low-dose XPCI to small-animal imaging one step further.